\documentclass[aps,prl,groupedaddress,twocolumn,10pt,reprint,superscriptaddress, longbibliography,floatfix]{revtex4-2}
\usepackage{epsfig,dsfont,amssymb,amsmath,amsthm,amsfonts,amsbsy,mathrsfs}
\usepackage{lipsum}
\usepackage{color}
\usepackage{graphicx}
\usepackage{amsmath}
\usepackage{amssymb}
\usepackage{subfigure}
\usepackage{float}
\usepackage{hyperref}
\usepackage{verbatim}
\usepackage{soul}
\usepackage{gensymb}
\usepackage{xeCJK}
\usepackage{comment}
\setstcolor{red}

\renewcommand{\eqref}[1]{\textrm{Eq.}(\ref{#1})}

\newcommand {\be}{\begin{equation}}
\newcommand {\ee}{\end{equation}}

\begin{document}
\title{Optimal Discretization in Hour-Glass Molecular Clocks Driven by Oscillating Free Energy}

\author{Zhuangcheng Zhen}
\affiliation{
School of Physics, Peking University, Beijing, 100086, China 
}
\author{Kaiyue Shi}
\affiliation{
School of Physics, Peking University, Beijing, 100086, China 
}
\affiliation{
Present address: Lewis-Sigler Institute for Integrative Genomics, Princeton University, Princeton, NJ 08544, USA
}
\author{Qi Ouyang}
\affiliation{
Institute for Advanced Study in Physics, School of Physics, Zhejiang University, Hangzhou 310058, China
}
\affiliation{
Center for Quantitative Biology, AAIC, Peking University, Beijing 100871, China
}
\author{Yuansheng Cao}
\affiliation{
Department of Physics, Tsinghua University, Beijing, 100084, China 
}

\begin{abstract} 
Hour-glass clocks do not free-run; they keep time by riding an external rhythm. Motivated by the primordial KaiBC system in cyanobacteria, we study a driven, finite-state molecular clock that advances through a small number of biochemical states under an intrinsic driving energy and a rotating energy landscape set by day–night metabolism. In the continuum limit, coherence is maximized at a resonant operating point where the intrinsic drift matches the driving frequency. In realistic clocks with a finite number of states, discreteness matters: as the rotating landscape sweeps over a lattice of states, it generates a small and high frequency vibration of the collective phase that makes timing inaccurate. Combining the resonant cost with this discreteness penalty yields a trade-off in the number of states: few states are energetically cheap but noisy; many states are precise but costly. The optimum lies at moderate discretization (typically five to fifteen states) and an environmental coupling that is strong enough for responsiveness yet weak enough to avoid large discrete-state vibrations. These design rules rationalize why KaiC’s hexameric architecture falls near the predicted optimum and suggest a general principle for hour-glass clocks across organisms.
\end{abstract}
\maketitle


\textbf{Introduction}.
Living cells use circadian clocks to track earth’s 24-h rotation and coordinate physiology~\cite{CircandianReview2003,mohawk2012central,2005ClockReview, 2020ClockReview}. Conceptually, timekeeping systems fall into two classes. Autonomous clocks generate self-sustained oscillations with a fixed intrinsic period—e.g., transcription–translation feedback loops in eukaryotic cells~\cite{1990PNASfeedback,review2017MammalsCircadian,1998CLOCKBMAL1,ko2006molecular},  the KaiABC oscillator in cyanobacteria~\cite{KaiCInVitro2005,KaiCOrigin1998,rust2007ordered} and neuron activities~\cite{2011BloodPrx,NeuronsFiring2014,NeuronsFiring2015}. Hour-glass clocks, by contrast, do not oscillate on their own: they progress through states at a set pace but complete (or “flip”) their cycle only under an external drive. One example is the primordial KaiBC system, where KaiA is absent and KaiC phosphorylation is paced by daily metabolic rhythms~\cite{KaiCEvolution2003,KaiBCLD2023,KaiA3B3C32024,anKaiCLD99Glow2025,KaiCEvo2025}; here the clock tracks time because the cellular free-energy landscape is periodically driven by day–night metabolism signal (e.g., oscillation in ATP/ADP ratio~\cite{2013RustATPADP,2017PCB,KaiBCLD2023}). These driven clocks raise a design question central to their functions: how should a molecular clock be structured to remain sensitive to a daily signal while keeping its timing coherent in an energetically efficient way~\cite{UdoTUR2015,cao2015free,Shuffling2020,2019KaiCTUR,barato2017coherence}, without relying on self-sustained dynamics\cite{UdoTURClock2016,UdoTimeTUR2020,RustIENoiseELife2018,emberly2006hourglass,gonze2005spontaneous}?

We address this question with a minimal, experimentally inspired model of a driven stochastic clock. The clock advances through a finite set of biochemical states on a ring; an intrinsic energetic tilt biases forward and backward steps, and a rotating potential encodes the daily metabolic drive (with period $T$). In the continuum limit, a simple resonance condition between the intrinsic drift and the external drive fixes the operating point and yields a dissipation that is essentially independent of the drive amplitude at resonance. In realistic finite-state clocks (state number $N$), discreteness matters: as the rotating drive sweeps across the lattice of phases, it induces collective phase vibration with period $T/N$. This discreteness-induced modulation inflates phase variance (timing noise), biases the average advancement rate, and can compromise phase-sensitive control. These facts immediately suggests two design levers—the state number $N$ and the drive amplitude —whose effects must be balanced.

Our central result is that, by balancing the dissipation at the resonance and collective phase vibration., we found the optimum occurs at moderate discretization (typically $N\sim 5-15$), and a sufficiently large drive amplitude that maximizes responsiveness without exciting large discrete-state vibration. These principles rationalize why primordial KaiBC-type hour-glass clocks, entrained by metabolism, would favor a handful of effective states rather than extreme coarse- or fine-graining. The desgin principles for how timing precision should depend on state number and drive strength could be extended to other hour-glass clocks.

\begin{figure}[t]\centering
\includegraphics[width=1\linewidth]{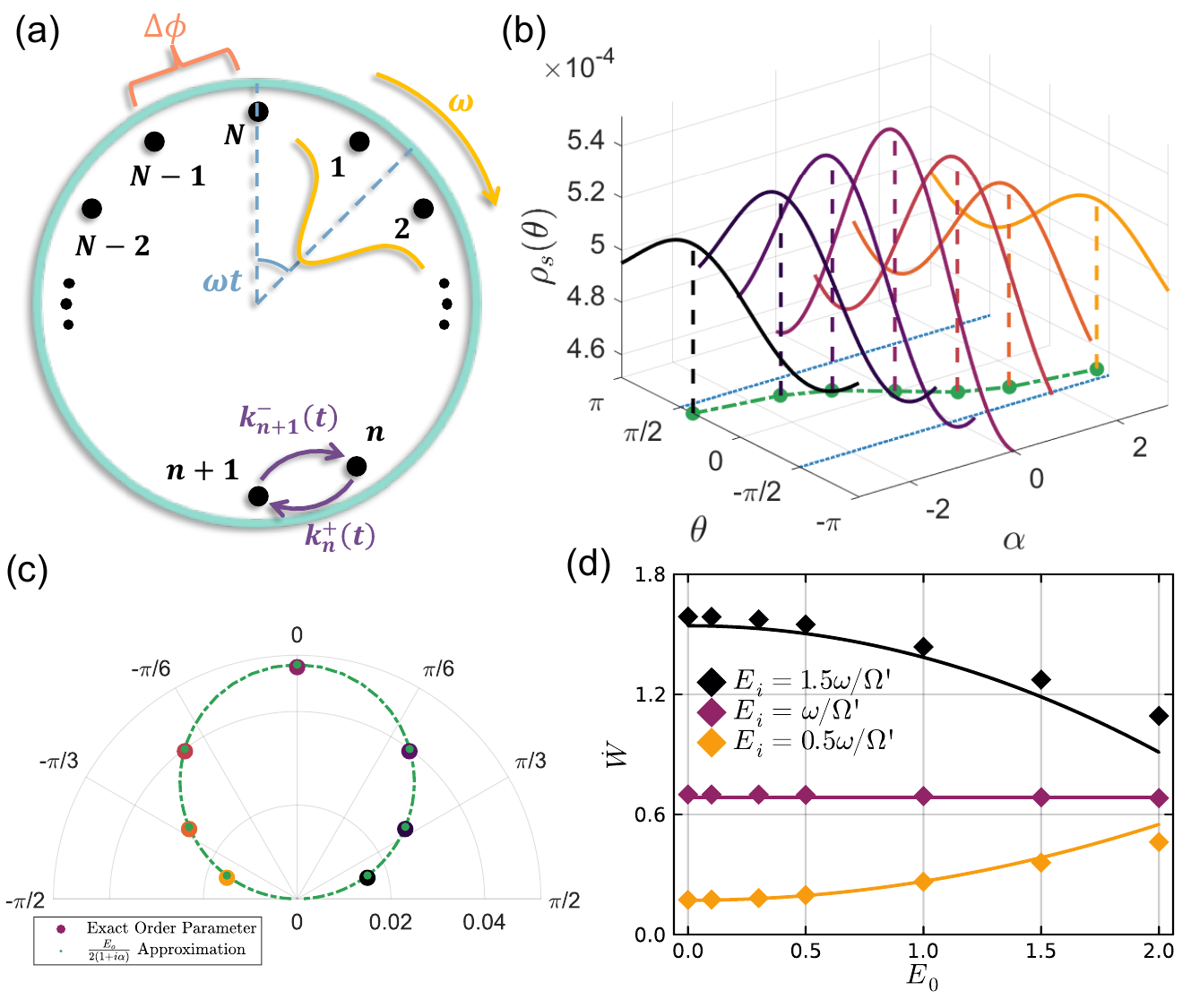}\\
\caption{Schematic illustration of the model and the resonant condition. (a) The system consists of $N$ states, representing the phase $n2\pi/N$, arranged on a ring with periodic boundary conditions. The forward and backward reactions are driven by both an intrinsic force and a time-dependent external potential. (b) Steady-state distribution in the rotating frame $\theta=\phi-\omega t$ under the continuous limit. The figure shows the distribution for different values of the parameter $\alpha=(\omega-\Omega^{'}E_i)/\Omega^{'}$. The distribution is narrowest at resonance ($\alpha=0$). The green dashed line represents the approximate theoretical prediction. (c) The order parameter for different values of $\alpha$, plotted on the complex plane. (d) Analytical and numerical results for the dissipation as a function of $E_o$ for various intrinsic driving strengths $E_i$ (with $N=500$). As $E_o$ increases, the dissipation for different $E_i$ values converges to the dissipation at resonance.}\label{fig1}
\end{figure}

\textbf{The model}.
We consider a stochastic ``hour-glass" molecular clock (i.e. Poisson clock) represented as a ring of $N$ discrete states, labeled $n=1,2,\cdots,N$ (Fig.~\ref{fig1}(a)). Each state corresponds to a phase angle $\phi_n=2\pi n/N$, with periodic boundary conditions $\phi_{N+1}=\phi_1$. Transitions occur between nearest neighbors, with rates $k_n^{\pm}$ describing the forward $(n\to n+1)$ and backward $(n\to n-1)$ jumps. These rates are modulated by both an intrinsic energetic bias and a periodic external signal.

The intrinsic drive assigns an energy different $E_i\Delta\phi$ between two adjacent states, with $\Delta\phi=2\pi/N$. Completing one full cycle consumes free energy $2\pi E_i$. In isolation, the mean drift rate is $(k_i^+-k_i^-)\Delta\phi$. However, without external synchronization, stochastic diffusion of the phase rapidly destroys coherence~\cite{kuramoto1975formation,2005kuramotoReview,cao2015free,UdoTURClock2016}, leading to a uniform distribution of all phases. To model entrainment by a metabolic rhythm, we introduce a time-dependent external potential
\[
U(\phi,t)=-E_o\cos(\phi-\omega t),
\]
which rotates at frequency $\omega$. Physically, this represents oscillations in the free-energy supply (e.g. ATP/ADP ratio) imposed by day–night cycling. The transition rates are then chosen to satisfy local detailed balance,
\begin{equation}\label{db}
\ln{\frac{k_n^+(t)}{k_{n+1}^-(t)}} = E_i \Delta\phi + E_o[ \cos{(\phi_{n+1} - \omega t)} - \cos{(\phi_{n} - \omega t)}],
\end{equation}
where we take the convention  $k_BT=1$. Without loss of generality, the forward and backward rates can be written as
\begin{align*}
k_n^+ &=\Omega\exp[IE_i\Delta\phi+O\Delta U_n(t)],\\
 k_{n+1}^-&=\Omega\exp[-(1-I)E_i\Delta\phi-(1-O)\Delta U_n(t)],
\end{align*}
with $\Delta U_n(t)=E_o[\cos(\phi_{n+1}-\omega t)-\cos(\phi_n-\omega t)]$. Unless stated otherwise, we adopt a convention $I=1,\;O=1/2$. Other choices of transition rates do not change the main conclusion of this study.

The clock's state distribution $P(\phi_n,t)$ is governed by the master equation
\begin{equation}\label{master eq}
\begin{split}
\frac{d P(\phi_n, t)}{d t} = &\ k_{n-1}^{+}(t)P(\phi_{n-1}, t) + k_{n+1}^{-}(t)P(\phi_{n+1}, t) \\ &- [k_{n}^{-}(t) + k_{n}^{+}(t)]P(\phi_n, t).
\end{split}
\end{equation}
To analyze the collective behavior, we define the order parameter
\begin{equation}\label{order para}
Re^{i\Psi}=\sum_{n=1}^NP(\phi_n,t)e^{i(\phi_n-\omega t)},
\end{equation}
with the collective amplitude $R\in [0,1]$ and the collective phase $\Psi$. At steady-state, $R$ measures the coherence of the ensemble, while vibration of $\Psi$ quantify the uncertainty in timing. The energetic cost is characterized by the dissipation rate
\begin{equation}
\dot{W}=\sum_n(J_n^+-J_{n+1}^-)\ln\frac{J_n^+}{J_{n+1}^-}
\end{equation}
where the forward and backward fluxes are defined as $J_n^+=k_n^+P(\phi_n,t)$ and $J_n^-=k_n^-P(\phi_n,t)$.

\begin{figure*}[t]\centering
\includegraphics[width=1\linewidth]{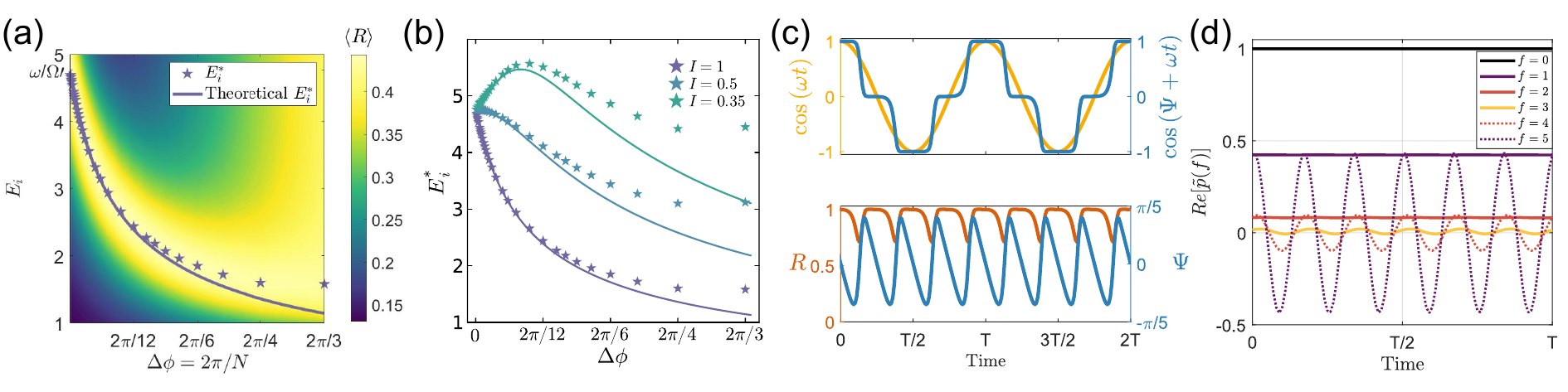}\\
\caption{Dissipation and phase vibration of the hour-glass clock. (a) Time-averaged collective amplitude $\left<R\right>$ as a function of intrinsic driving $E_i$ and step size $\Delta\phi$. The asterisks mark the optimal driving $E_i^{\ast}$ that yields the maximum amplitude for a fixed number of states $N$. The solid curves represent the theoretical predictions for $E_i^{\ast}$. Since $R$ exhibits small vibration over time, we plot its time-averaged value, $\left<R\right>$. (b) Relation between the optimal intrinsic driving $E_i^{\ast}$ and the step size $\Delta\phi$ for different partitions $I$. The case $I=1$ yields the minimum $E_i^{\ast}$. Parameters used:$\Omega'=0.055, \omega=2\pi/24$ (c) Time evolution of the collective amplitude $R$ and collective phase $\Psi$ for a small number of states $N=4$ and a large external potential $E_o=10$. Both the amplitude and phase vibrate with a period of $T/N$. When $E_o/N$ is large, the total phase $\omega t +\Psi$ is strongly pinned to the discrete phases $n2\pi/N$ of the clock, leading to large vibration in $\Psi$. (d) Time evolution of different Fourier modes in the rotating reference frame $\tilde{p}(f)$ for $N=6$. The amplitude of mode $n$ is identical to that of mode $N-n$. Modes closer to $N/2$ exhibit a smaller amplitude.}\label{fig2}
\end{figure*}

\textbf{The continuous limit and the resonance condition.}
In the continuous phase limit $N\to\infty$ ($\Delta\phi\to 0$), the master equation reduces to a Fokker-Planck equation in the rotating frame $\theta=\phi-\omega t$ (see Supplemental Material (SM) for details)
\begin{equation}\label{FP}
\frac{\partial\rho(\theta,t)}{\partial t} = \frac{\partial}{\partial\theta}\left[\omega-\Omega'\left(E_o\sin{\theta}-E_i+\frac{\partial}{\partial\theta}\right)\right]\rho(\theta),
\end{equation}
with $\Omega'=\Omega\Delta\phi^2$. $\rho(\theta,t)$ is the probability density of phase $\phi$ in the rotating frame. The steady-state solution $\rho_s(\theta)$ in the rotating reference frame is (see SM for details)
\begin{equation}\label{FP_ss}
\rho_s(\theta) = \frac{1}{Z}\left[\left(e^{2\pi\alpha}-1\right)\frac{I(\theta)}{I(2\pi)}+1\right]e^{-\alpha\theta+E_o\cos\theta},
\end{equation}
where $\alpha=(\omega-\Omega'E_i)/\Omega'$, $I(y) = \int_0^y e^{\alpha x-E_o\cos x}dx$, and $Z$ is the normalization constant (Fig.~\ref{fig1}(b)). The order parameter now becomes $Re^{i\Psi}=\int_0^{2\pi}\rho_s(\theta)e^{i\theta}d\theta$. For weak driving ($E_o\ll 1$), it follows the compact form (Fig.~\ref{fig1}(c))
\begin{equation}\label{op_ss}
Re^{i\Psi} = \frac{E_o}{2\left( 1 + i\alpha \right)},
\end{equation}
from which we obtain $R=E_o/\sqrt{1+\alpha^2}$. Hence coherence (represented by $R$) is maximal when $\alpha=0$, i.e. $\omega=\Omega' E_i$. The intrinsic frequency of the clock is $(k^+-k^-)\Delta\phi\approx \Omega E_i\Delta\phi^2=\Omega'E_i$. This indicates that the optimal coherence is achieved at resonance $\omega=\Omega'E_i$. The intrinsic driving free energy that maximizes coherence is
\begin{equation}\label{op_drive}
E_i^\ast=\omega/\Omega'.
\end{equation}


Furthermore, in the continuous limit, the dissipation rate can be calculated from the expression $\dot{W}=\int_0^{2\pi} J(\phi)^2/\left(\Omega'\rho(\phi)\right)d\phi$. Expanding $\dot{W}$ in powers of $E_o$ yields:
\begin{equation}\label{dp_ss}
\dot{W}=\Omega'\left[{E_i}^2 + \frac{2\alpha E_i + \alpha^2}{2\left(\alpha^2 +1 \right)}{E_o}^2 \right] + O\left({E_o}^3 \right)
\end{equation}
Figure~\ref{fig1}(d) compares the theoretical and numerical results. In this limit, the dissipation is dominated by $E_i$, while the $E_o$-dependent correction vanishes at resonance ($\alpha=0$). This behavior has a clear physical interpretation: when the rotating potential is sufficiently strong, the probability distribution becomes tightly localized near the instantaneous minimum of the external landscape, $\phi\approx\omega t$, largely independent of the intrinsic drift. As a result, the cycle-averaged energetic cost is set primarily by the resonant drive, with only higher-order contributions from $E_o$ (see SM for detail discussions). Evaluated at the coherence-optimal drive $E_i^*=\omega/\Omega'$, the dissipation at resonant becomes
\begin{equation}\label{dp_a0}
\dot{W}_{\text{res}}=\Omega'{E_i^*}^2= \frac{\omega^2}{\Omega'}=\omega E_i^\ast,
\end{equation}
independent of $E_o$, and which is exactly the intrinsic drive multiplied by the environmental frequency.


\begin{figure*}[t]\centering
\includegraphics[width=1\linewidth]{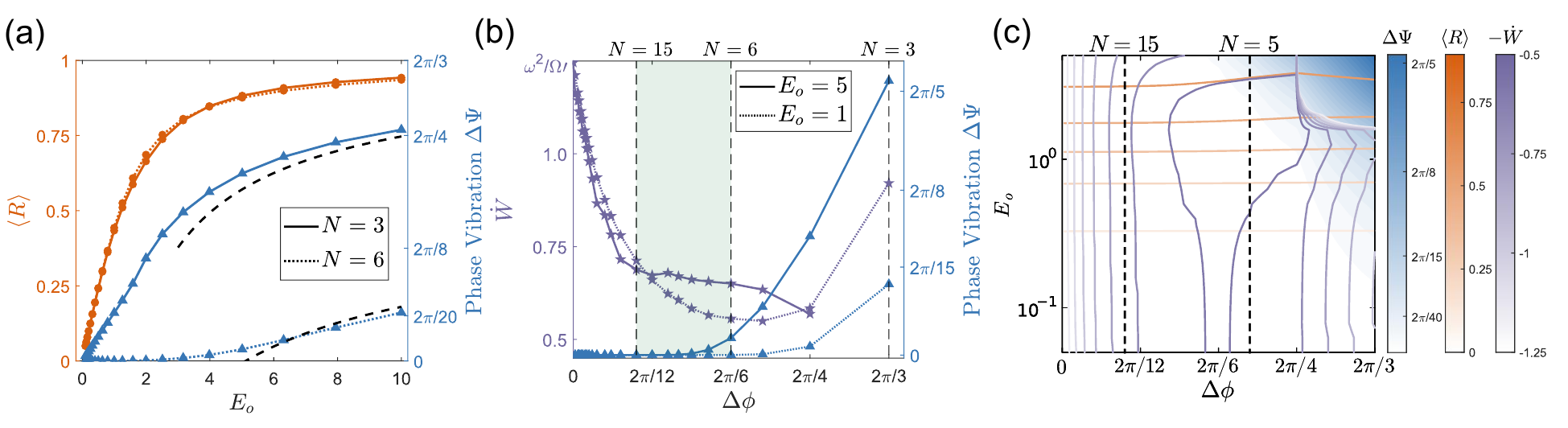}\\
\caption{Trade-off between driving cost and phase vibration controlled by the external signal strength $E_o$ and the clock's resolution $\Delta\phi$. Simulation results are numerically obtained at the resonant $E_i$ for each set of parameters ($\Omega'=0.055, \omega=2\pi/24$). (a) Both the collective amplitude $\left<R\right>$ and phase vibration increase with $E_o$. A larger number of states $N$ (higher resolution) is required to suppress this vibration. The black dashed line represents the approximate theoretical value. (b) However, a larger $N$ leads to higher dissipation. Thus, $N$ must be chosen within an optimal range to suppress phase vibration in an energy-efficient way. (c) Contour plots of phase vibration $\Delta\Psi$, collective amplitude $\left<R\right>$, and negative dissipation rate $-\dot{W}$ in the $E_o$-$\Delta\phi$ parameter space. Balancing the dissipation and phase vibration, an optimal state number is $N=5\textendash15$. Within this range, maximizing $E_o$ emerges as a viable optimization strategy.}\label{fig3}
\end{figure*}

\textbf{The discrete case}.
In real biological systems, molecular clocks consist of a finite number of states. For instance, the KaiBC system can be modeled with $N=14$ states (7 states for the phosphorylated conformation and 7 states for the dephosphorylated conformation~\cite{2002hexamer,2007DAB,2008sasai}). We now turn to the case of a discrete state space. To facilitate comparison across systems with different $N$, we hold the parameter $\Omega' = \Omega \Delta\phi^2$ constant, which is proportional to the phase diffusion on a ring with finite states \cite{cao2015free, Shuffling2020}. Applying a discrete Fourier transform $\hat{p}(f) = \sum_{n=0}^{N-1} P(\phi_n,t) e^{i\frac{2\pi f}{N}n}, \quad (f=0,1,...,N-1)$, we define the Fourier modes in the rotating reference frame as
\[
\tilde{p}(f) = \hat{p}(f)e^{-if\omega t}.
\]
The order parameter defined in Eq.~(\ref{order para}) is exactly the first mode $f=1$ in frequency domain:
\[
\tilde{p}(1) = \sum_{n=0}^{N-1} p(n) e^{i\left(\frac{2\pi n}{N}-\omega t\right)} = Re^{i\Psi}. 
\]
Expanding the master equations to first order of $E_o\Delta\phi$ and neglecting the contributions from $|\tilde{p}(2)|$ (Fig~\ref{fig2}(d))(see SM for details), 
we can obtain an approximate solution for the order parameter at steady-state:
\[
\tilde{p}_s(1) \approx \frac{E_o}{2+i4\dfrac{\omega\Delta\phi-\Omega'(e^{E_i\Delta\phi}-1)}{\Omega'\Delta\phi(e^{E_i\Delta\phi}+1)}}.
\]
The maximal coherence condition is the vanishing of the imaginary part of its denominator, giving the resonance condition at discrete case:
\begin{equation}\label{eq20}
\omega\Delta\phi = \Omega'(e^{E_i^\ast \Delta\phi}-1).
\end{equation}
and the optimal intrinsic drive:
\begin{equation}\label{eq19_v2}
E_i^* = \frac{1}{\Delta\phi}\ln\left(1+\frac{\omega\Delta\phi}{\Omega'}\right) = \frac{N}{2\pi}\ln\left(1+\frac{2\pi\omega}{N\Omega'}\right).
\end{equation}
As $N$ increases, $E_i^*$ increases, so does the dissipation rate. In the limit $N\to\infty$, the expansion of the logarithm recovers the continuous-limit result, $E_i^\ast=\omega/\Omega'$, as shown in Fig.~\ref{fig2}(a).

More generally, different intrinsic partitions $I$ (how the intrinsic drive is split between lowering forward v.s. raising reverse barriers) modify the resonance condition to (see SM for details)
\begin{equation}\label{reso_IO}
e^{IE_i^*\Delta\phi} - e^{-(1-I)E_i^*\Delta\phi} = \frac{\omega\Delta\phi}{\Omega'}.
\end{equation}
For any partition with $I<1$, $E_i^*(I<1)\geq E_i^*(I=1)$ (Fig.~\ref{fig2}(b)). This implies that allocating the intrinsic drive to decrease the backward rate, rather than increase the forward rate, is less efficient. To minimize dissipation at resonance, the driving energy should primarily enhance the forward transitions. Similar results can be obtained for different partitions on the external drive $O$ (see SM for details).

While the preceding analysis relied on a steady-state approximation, the order parameter is in fact not constant but exhibits weak vibration $\Delta\Psi$ (Fig.~\ref{fig2}(c)). Mechanistically, the discrete phase lattice cannot track the continuously rotating drive smoothly. Because the local probability is weighted by a nonlinear factor  $\exp(-U(\phi_n,t))$ with $U(\phi_n,t)=-E_o\cos(\phi_n-\omega t)$, contributions from the two states flanking $\omega t$ do not cancel exactly as the drive sweeps across $[\phi_n,\phi_{n+1}]$, thus producing a small oscillation that repeats $N$ times per cycle. In the large-$E_o$ regime the vibration amplitude approaches $\Delta\phi$; for large but finite $E_o$, we have the phase vibration $\Delta\Psi=\Psi_{\max}-\Psi_{\min}$(see SM for details):
\begin{equation}\label{eqNutat}
\Delta\Psi \approx \frac{2\pi}{N} - \frac{1 + \ln\left(\frac{4\pi E_o \sin(\pi/N)}{N}\right)}{E_o \sin(\pi/N)}.
\end{equation}
From the perspective of the mode equation, the mode $f=1$ couples to its conjugate $f=N-1$,
\[
\tilde{p}(N-1) = \tilde{p}(1)^*e^{-iN\omega t},
\]
so feedback at frequency $N\omega$ drives the $T/N$ vibration. This coupling first appears at second order in the expansion in $E_o\Delta\phi$ (see SM for details), explaining why the vibration grows with $E_o/N$ and vanishes in the limit of $N\to\infty$.

\textbf{Optimal discretization}.
 Our analysis reveals two fundamental trade-offs governed by the number of states $N$ and the drive amplitude $E_o$. As shown in Fig.~\ref{fig3}(a), increasing $E_o$ strengthens entrainment in the continuum limit, but in the discrete clock it also amplifies the finite-$N$ vibration, thereby reducing timing accuracy. The collective amplitude $\langle R\rangle$ is nearly independent of $N$. By contrast, increasing $N$ suppresses the discrete-state vibration, but it also raises the energetic cost ($E_i^*$) at the resonance point (Fig.~\ref{fig3}(b)) (The increase in $\dot{W}$ at low $N$ is explained in SI). Few states are cheap but noisy; many states are precise but costly. Balancing these effects, the clock should operate with a moderate number of states (e.g., $N \approx 5\textendash15$), while using $E_o$ large enough to ensure responsiveness but not so large that discrete-state vibrations dominate (Fig.~\ref{fig3}(c)).

\textbf{Discussion}.
Inspired by the primordial KaiBC system, we developed a toy model for an hour-glass molecular clock whose energy landscape is driven by external environmental signals. The model is characterized by an intrinsic driving energy $E_i$ and an external environmental coupling strength $E_o$. The intrinsic drive provides an intrinsic frequency. When this frequency matches the external  signal's frequency $\omega$, the system achieves maximal oscillation amplitude, and its dissipation is primarily governed by the resonant drive $E_i^*$. We find that $E_i^*$ decreases as the number of states $N$ decreases, suggesting lower dissipation for smaller systems. However, a small $N$ introduces significant vibrations of the collective phase due to state space discreteness, which reduces timekeeping precision. Effective timekeeping also demands a large $E_o$ for a high collective amplitude, which also increases the phase vibration. Therefore, an optimal design principle involves a trade-off: selecting an appropriate $N$ to minimize dissipation and phase vibration while maximizing environmental sensitivity through $E_o$. 

In cyanobacteria, circadian proteins act as a timing gate that modulates downstream transcriptional programs, coordinating processes such as cell division, photosynthesis and enzyme sequestration~\cite{Oshea2013TTFL, AndyLiWang2021KaiBScience, susan2015review, Locke2018DivisionClock}. For survival, this gate must switch rapidly and reliably. Our results show that discrete-state phase vibration (Fig.~\ref{fig2}(c)) introduces phase ambiguity, reducing the accuracy of such phase-sensitive control. Notably, the KaiC hexamer naturally realizes 14 effective states, which falls within our predicted optimal range for balancing cost and precision. Moreover, ring-shaped hexameric ATPases recur as clock components across diverse organisms~\cite{ju2020chemical,RUVBL2025,2021ruvbl1Review}, suggesting that oligomeric state—and the associated effective discretization—has been evolutionarily selected to optimize timekeeping. We therefore propose that choosing an appropriate state number to minimize dissipation and vibration while maintaining responsiveness is a general design principle of hour-glass molecular clocks.

\textbf{Acknowledgment}. The work by Z. Z. and Y. C. are supported by the National Key Research and Development Program of China (Grant No.2024YFA0919600). 

\bibliography{refer.bib}

\end{document}


\maketitle


This Supplemental Material mainly details the mathematical derivations of the formalisms presented in the main text, and briefly extends the discussion on detail problems. First, we derive the exact solution using the Fokker-Planck equation in the continuous state-space limit. For weak coupling strength $E_o$, we provide a perturbative expansion to approximate the order parameter and energy dissipation. For systems with a finite number of states, we employ the discrete Fourier transform to analyze the order parameter and the periodic ``phase vibration'' induced by mode coupling. Additionally, we examine the effects of asymmetric external potential partitioning ($O \neq 0.5$), extending the analysis beyond the symmetric case discussed in the main text. Finally, we address the anomalous dissipation observed at $N=3$ [Fig.~3(b)] and provide a physical interpretation for the convergence of dissipation toward the resonant value in the large $E_o$ limit.

\tableofcontents

\section{Fokker-Planck Equation}

In the continuous state-space limit ($\Delta\phi\to0$), a Kramers-Moyal expansion reduces the master equation to a Fokker-Planck equation:
\begin{equation}\label{eq1}
\begin{aligned}
\frac{\partial P(\phi_n, t)}{\partial t} &= k_{n-1}^{+}(t)P(\phi_{n-1}, t) + k_{n+1}^{-}(t)P(\phi_{n+1}, t) - [k_n^{-}(t) + k_n^{+}(t)]P(\phi_n, t) \\
&\approx \left(k_n^{+}(t) - \frac{\partial k_n^{+}(t)}{\partial\phi_n}\Delta\phi + \frac{1}{2}\frac{\partial^2 k_n^{+}(t)}{\partial\phi_n^2}\Delta\phi^2\right)\left(P(\phi_n, t) - \frac{\partial P(\phi_n, t)}{\partial\phi_n}\Delta\phi + \frac{1}{2}\frac{\partial^2 P(\phi_n, t)}{\partial\phi_n^2}\Delta\phi^2\right) \\
&\quad + \left(k_n^{-}(t) + \frac{\partial k_n^{-}(t)}{\partial\phi_n}\Delta\phi + \frac{1}{2}\frac{\partial^2 k_n^{-}(t)}{\partial\phi_n^2}\Delta\phi^2\right)\left(P(\phi_n, t) + \frac{\partial P(\phi_n, t)}{\partial\phi_n}\Delta\phi + \frac{1}{2}\frac{\partial^2 P(\phi_n, t)}{\partial\phi_n^2}\Delta\phi^2\right) \\
&\quad - [k_n^{-}(t) + k_n^{+}(t)]P(\phi_n, t) \\
&\approx \left(\frac{\partial k_n^{-}(t)}{\partial\phi_n}-\frac{\partial k_n^{+}(t)}{\partial\phi_n}\right)P(\phi_n, t)\Delta\phi + (k_n^{-}(t)-k_n^{+}(t))\frac{\partial P(\phi_n, t)}{\partial\phi_n}\Delta\phi \\
&\quad + (k_n^{-}(t)+k_n^{+}(t))\frac{1}{2}\frac{\partial^2 P(\phi_n, t)}{\partial\phi_n^2}\Delta\phi^2 \\
&= \Omega\left[E_o\sin(\phi-\omega t)P(\phi_n, t) + (E_o\sin(\phi-\omega t)-E_i)\frac{\partial P(\phi_n, t)}{\partial\phi_n}+\frac{\partial^2 P(\phi_n, t)}{\partial\phi_n^2}\right]\Delta\phi^2 \\
&= \Omega\frac{\partial}{\partial\phi}\left[E_o\sin(\phi-\omega t)-E_i+\frac{\partial}{\partial\phi}\right]P(\phi_n, t)\Delta\phi^2.
\end{aligned}
\end{equation}

To ensure the invariance of the equation properties and distribution shape regardless of discretization, the reaction rates are rescaled by $\Omega^{'} =\Omega{\Delta \phi}^2$, yielding:
\begin{equation}\label{fpeq}
\frac{\partial\rho(\phi)}{\partial t} = \frac{d}{d\phi}\left[\Omega'\left(E_o\sin(\phi-\omega t)-E_i+\frac{d}{d\phi}\right)\right]\rho(\phi).
\end{equation}

To eliminate the explicit time dependence, we transform to the rotating coordinate frame $\theta=\phi-\omega t$:
\begin{equation}\label{eq2}
\frac{\partial\rho(\theta)}{\partial t} = \frac{d}{d\theta}\left[\omega+\Omega'\left(E_o\sin\theta-E_i+\frac{d}{d\theta}\right)\right]\rho(\theta).
\end{equation}


\section{Derivation of the Steady-State Solution}

The steady-state condition $\frac{\partial \rho_s}{\partial t} = 0$ implies a constant probability current $J_s$:
\begin{equation}\label{Jss}
    \Omega' \frac{d\rho_s}{d\theta} + \left[\omega + \Omega' (E_o \sin\theta - E_i)\right]\rho_s = J_s.
\end{equation}

Letting $\alpha = (\omega - \Omega' E_i) / \Omega'$, the equation becomes a first-order linear ODE:
\begin{equation}
    \frac{d\rho_s}{d\theta} + (\alpha + E_o \sin\theta)\rho_s = \frac{J_s}{\Omega'}.
\end{equation}

The general solution is obtained using the integrating factor $\mu(\theta) = \exp(\alpha\theta - E_o \cos\theta)$:
\begin{equation}
    \rho_s(\theta) = \mu(\theta)^{-1} \left[ C_1 + \frac{J_s}{\Omega'} \int_0^\theta \mu(x) dx \right].
\end{equation}

Defining $I(y) = \int_0^y e^{\alpha x - E_o \cos x} dx = \int_0^y \mu(x) dx$ and setting $C_1 = \rho_s(0)e^{-E_o}$, we have:
\begin{equation}
    \rho_s(\theta) = e^{-\alpha\theta + E_o \cos\theta} \left[ C_1 + \frac{J_s}{\Omega'} I(\theta) \right].
\end{equation}

Applying the periodic boundary condition $\rho_s(2\pi) = \rho_s(0)$ allows us to solve for $J_s$:
\begin{equation}
    C_1 e^{E_o} = e^{-2\pi\alpha + E_o} \left[ C_1 + \frac{J_s}{\Omega'} I(2\pi) \right] \implies \frac{J_s}{\Omega'} = C_1 \frac{e^{2\pi\alpha} - 1}{I(2\pi)}.
\end{equation}

Substituting this back into the general solution with the normalization constant $Z = 1/C_1$, we obtain:
\begin{equation}\label{rhoSS}
    \rho_s(\theta) = \frac{1}{Z}\left[(e^{2\pi\alpha} - 1)\frac{I(\theta)}{I(2\pi)} + 1\right]e^{-\alpha\theta + E_o \cos\theta}.
\end{equation}


\section{First-order Expansion of the Steady-State Solution and Order Parameter}

The exact steady-state solution involves integrals that complicate the calculation of the order parameter. For small coupling strength $E_o$, we perform a perturbative expansion to approximate the steady-state solution. Expanding the exponential term:
\begin{equation}
e^{E_o\cos\theta} = 1 + E_o\cos\theta + O(E_o^2).
\end{equation}

The integral $I(\theta)$ is approximated as:
\begin{equation}
I(\theta) = \int_0^{\theta} e^{\alpha x - E_o\cos x}\,dx
= I_0(\theta) - E_o I_1(\theta) + O(E_o^2),
\end{equation}
where
\begin{equation}
I_0(\theta) = \int_0^{\theta} e^{\alpha x}\,dx = \frac{e^{\alpha\theta}-1}{\alpha},
\end{equation}
\begin{equation}
I_1(\theta) = \int_0^{\theta} e^{\alpha x}\cos x\,dx = \frac{e^{\alpha\theta}(\alpha\cos\theta+\sin\theta) - \alpha}{1+\alpha^2}.
\end{equation}

Substituting these expansions into Eq.~(\ref{rhoSS}) yields the first-order approximation:
\begin{equation}
\rho_s(\theta)
=\frac{1}{Z}\big[1+E_o F(\theta)\big]+O(E_o^2),
\qquad
F(\theta)=\frac{\alpha^2+\cos\theta-\alpha\sin\theta}{1+\alpha^2}.
\end{equation}

To determine the normalization constant $Z$, we calculate the mean of $F$:
\begin{equation}
\overline{F} = \frac{1}{2\pi}\int_0^{2\pi}F(\theta)\,d\theta = \frac{\alpha^2}{1+\alpha^2}.
\end{equation}
Normalization requires $\int_0^{2\pi}\rho_s(\theta)\,d\theta=\frac{2\pi}{Z}(1+E_o\overline{F})=1$. Correcting the prefactor up to $O(E_o)$ gives:
\begin{equation}
\frac{1}{Z} = \frac{1}{2\pi}\big(1 - E_o\overline{F}\big) + O(E_o^2).
\end{equation}

Thus, the normalized first-order expansion is:
\begin{equation}\label{ss_solu}
\rho_s(\theta)
= \frac{1}{2\pi}
\left[
1 + E_o\big(F(\theta)-\overline{F}\big)
\right] + O(E_o^2)
= \frac{1}{2\pi}
\left[
1 + \frac{E_o}{1+\alpha^2}\big(\cos\theta - \alpha\sin\theta\big)
\right] + O(E_o^2).
\end{equation}

Consequently, the order parameter is derived as:
\begin{equation}
R e^{i\psi}=\int_0^{2\pi}\rho_s(\theta)e^{i\theta}d\theta=\frac{E_o}{2(1+i\alpha)}+O(E_o^2).
\end{equation}


\section{Dissipation}

In the continuous limit, energy dissipation can be approximated as:
\begin{equation}\label{dissi}
\dot{W} = \int_0^{2\pi} \frac{J(\phi)^2}{\Omega'\rho(\phi)}\,d\phi= \int_0^{2\pi} \frac{J(\theta)^2}{\Omega'\rho(\theta)}\,d\theta.
\end{equation}
Note that $J(\theta)$ differs from the steady-state flux $J_s$ in Eq.~(\ref{Jss}). $J_s$ represents the constant probability current in the rotating frame, while $J(\theta)$ denotes the flux in the original frame, expressed in rotating coordinates. From Eq.~(\ref{fpeq}), we have:
\begin{equation}
J(\theta) = \Omega'\left(E_o\sin\theta-E_i+\frac{d}{d\theta}\right)\rho(\theta).
\end{equation}
Substituting this and Eq.~(\ref{ss_solu}) into Eq.~(\ref{dissi}) and expanding to second order in $E_o$. Most of the terms vanish upon integration as they are trigonometric. So we obtain:
\begin{equation}\label{dp_ss}
\dot{W}=\Omega'\left[{E_i}^2 + \frac{2\alpha E_i + \alpha^2}{2\left(\alpha^2 +1 \right)}{E_o}^2 \right] + O\left({E_o}^3 \right).
\end{equation}


\section{Fourier Transform of the ODEs}

For systems with a finite number of states, the Fokker-Planck approximation breaks down, and direct solution of the master equation remains difficult. We therefore transform the equations into Fourier space. As noted in the main text, the first Fourier mode in the rotating frame corresponds to the order parameter. The discrete Fourier transform is defined as:
\begin{equation}
\hat{p}(f) = \sum_{n=0}^{N-1} P(\phi_n,t) e^{i\frac{2\pi f}{N}n}, \quad (f=0,1,...,N-1).
\end{equation}
Taking the derivative of $\hat{p}(f)$:
\begin{equation}\label{eq3}
\begin{aligned}
\frac{d\hat{p}(f)}{dt} = &\sum_n e^{i\frac{2\pi f n}{N}} \frac{dp(n)}{dt} \\
= &\sum_n e^{i\frac{2\pi f n}{N}} [k^+(n-1)p(n-1) + k^-(n+1)p(n+1) - (k^+(n) + k^-(n))p(n)] \\
= &\sum_n -e^{i\frac{2\pi f n}{N}} p(n)(k^+(n)+k^-(n)) + \sum_n [ e^{i\frac{2\pi f}{N}n}e^{i\frac{2\pi f}{N}} k^+(n)p(n) + e^{i\frac{2\pi f }{N}n}e^{-i\frac{2\pi f}{N}} k^-(n)p(n)] \\
= &\sum_n -e^{i\frac{2\pi f n}{N}} p(n)(k^+(n)+k^-(n)) \\
&+ \sum_n e^{i\frac{2\pi f n}{N}} \left[k^+(n)p(n) \left[\cos\left(\frac{2\pi f}{N}\right) + i \sin\left(\frac{2\pi f}{N}\right)\right] + k^-(n)p(n) \left(\cos\left(\frac{2\pi f}{N}\right) - i \sin\left[\frac{2\pi f}{N}\right]\right)\right] \\
= &\ 2 \sin^2\left(\frac{\pi f}{N}\right) \sum_n e^{i\frac{2\pi f }{N}n} (k^+(n)p(n) + k^-(n)p(n)) \\
&+ i \sin\left(\frac{2\pi f}{N}\right) \sum_n e^{i\frac{2\pi f }{N}n} (k^+(n)p(n) - k^-(n)p(n)).
\end{aligned}
\end{equation}

Using the convolution theorem, we arrive at:
\begin{equation}\label{eq4}
\begin{aligned}
\frac{d\hat{p}(f)}{dt} = &-2\sin^2\left(\frac{\pi f}{N}\right)[(\hat{k}^{+}\ast\hat{p})(f) + (\hat{k}^{-}\ast\hat{p})(f)] \\
&+ i\sin\left(\frac{2\pi f}{N}\right)[(\hat{k}^{+}\ast\hat{p})(f) - (\hat{k}^{-}\ast\hat{p})(f)],
\end{aligned}
\end{equation}
where $(\hat{k}^{\pm} \ast \hat{p})(f) = \frac{1}{N}\sum_{l=0}^{N-1}\hat{k}^{\pm}(l)\cdot\hat{p}(f-l)$ is the circular convolution, and $\hat{k}^{\pm}(f)$ represents the transition rates in the frequency domain:
\begin{equation}
\hat{k}^{\pm}(f) = \sum_{n=0}^{N-1} e^{i\frac{2\pi f}{N}n} \cdot k^{\pm}(n) , \quad (f=0,1,...,N-1).
\end{equation}


\section{Derivation and Solution of the Mode Equations}

To facilitate the convolution, we Taylor expand the phase-dependent component of $k^{\pm}(n)$ to first order, as retaining the full exponential form renders calculation intractable:
\begin{equation}\label{eq5}
k^\pm(n) = \Omega e^{\pm I^{\pm}E_t \Delta\phi} \left[ 1 \mp O^\pm E_o \sin(\phi-\omega t)\Delta\phi \right].
\end{equation}

Here, we change the symbols for simplicity: $I^+=I,~I^-=1-I,~O^+=O,~O^-=1-O$. In Fourier space, this becomes:

\begin{equation}\label{eq6}
\hat{k}^\pm(f) = \Omega e^{\pm I^{\pm}E_t \Delta\phi} \left[ \delta_{f,0} \pm \delta_{f,1} O^\pm E_o \Delta\phi \frac{ie^{-i\omega t}}{2} \mp \delta_{f,N-1} O^\pm E_o \Delta\phi \frac{ie^{i\omega t}}{2} \right].
\end{equation}

The convolution yields:

\begin{equation}\label{eq7}
(\hat{k}^\pm * \hat{p})(f) = \Omega e^{\pm I^{\pm}E_t\Delta\phi} \left[ \hat{p}(f) \pm O^\pm \frac{iE_o\Delta\phi}{2} \left( e^{-i\omega t}\hat{p}(f+1) - e^{i\omega t}\hat{p}(f-1) \right) \right].
\end{equation}

Substituting this into Eq.~(\ref{eq4}) gives the mode evolution equation:

\begin{equation}\label{eq8}
\begin{aligned}
\frac{d\hat{p}(f)}{dt} =& -2\sin^2\left(\frac{\pi f}{N}\right) \Omega\hat{p}(f) (e^{{I}^{+}E_i\Delta\phi} + e^{-{I}^{-}E_i\Delta\phi}) \\
& + \sin\left(\frac{2\pi f}{N}\right) \Omega\frac{E_o\Delta\phi}{2}[e^{i\omega t}\hat{p}(f-1) - e^{-i\omega t}\hat{p}(f+1)]({O}^{+}e^{{I}^{+}E_i\Delta\phi} + {O}^{-}e^{-{I}^{-}E_i\Delta\phi}) \\
& + i\sin\left(\frac{2\pi f}{N}\right) \Omega\hat{p}(f)(e^{{I}^{+}E_i\Delta\phi} - e^{-{I}^{-}E_i\Delta\phi}) \\
& - i\sin^2\left(\frac{\pi f}{N}\right) \Omega E_o\Delta\phi[e^{i\omega t}\hat{p}(f-1) - e^{-i\omega t}\hat{p}(f+1)]({O}^{+}e^{{I}^{+}E_i\Delta\phi} - {O}^{-}e^{-{I}^{-}E_i\Delta\phi}).
\end{aligned}
\end{equation}

Applying the transformation $\tilde{p}(f) = \hat{p}(f)e^{-i\omega tf}$ eliminates the explicit time dependence:

\begin{equation}\label{eq9}
\begin{aligned}
\frac{d\tilde{p}(f)}{dt} =& -2\sin^2\left(\frac{\pi f}{N}\right) \Omega\tilde{p}(f) (e^{{I}^{+}E_i\Delta\phi} + e^{-{I}^{-}E_i\Delta\phi}) \\
& + \sin\left(\frac{2\pi f}{N}\right) \Omega\frac{E_o\Delta\phi}{2}[\tilde{p}(f-1) - \tilde{p}(f+1)]({O}^{+}e^{{I}^{+}E_i\Delta\phi} + {O}^{-}e^{-{I}^{-}E_i\Delta\phi}) \\
& + i\sin\left(\frac{2\pi f}{N}\right) \Omega\tilde{p}(f)(e^{{I}^{+}E_i\Delta\phi} - e^{-{I}^{-}E_i\Delta\phi}) - if\omega\tilde{p}(f) \\
& - i\sin^2\left(\frac{\pi f}{N}\right) \Omega E_o\Delta\phi[\tilde{p}(f-1) - \tilde{p}(f+1)]({O}^{+}e^{{I}^{+}E_i\Delta\phi} - {O}^{-}e^{-{I}^{-}E_i\Delta\phi}).
\end{aligned}
\end{equation}

Note that, the first-order expansion results in only nearest-neighbor interactions in the frequency domain.


\section{Approximate Solution of $\tilde{p}(1)$}

Given the normalization $\tilde{p}(0)=1$, and assuming $\tilde{p}(2)$ is negligible (being a higher-order term relative to the order parameter $\tilde{p}(1)$), we solve for the steady-state $\tilde{p}_s(1)$:

\begin{equation}\label{eq10}
\tilde{p}_s(1) \approx \frac{E_o({O}^{+}e^{{I}^{+}E_i\Delta\phi} + {O}^{-}e^{-{I}^{-}E_i\Delta\phi}) + i2E_o\Delta\phi({O}^{+}e^{{I}^{+}E_i\Delta\phi} - {O}^{-}e^{-{I}^{-}E_i\Delta\phi})}{\left(e^{{I}^{+}E_i\Delta\phi} + e^{-{I}^{-}E_i\Delta\phi}\right) + i2\left[\dfrac{\omega}{\Omega'}-\dfrac{1}{\Delta\phi}(e^{{I}^{+}E_i\Delta\phi} - e^{-{I}^{-}E_i\Delta\phi})\right]}.
\end{equation}

Neglecting the second term in the numerator (higher order), we obtain:

\begin{equation}\label{eq11}
\tilde{p}_s(1) \approx \frac{E_o({O}^{+}e^{{I}^{+}E_i\Delta\phi} + {O}^{-}e^{-{I}^{-}E_i\Delta\phi})}{\left(e^{{I}^{+}E_i\Delta\phi} + e^{-{I}^{-}E_i\Delta\phi}\right) + i2\left[\dfrac{\omega}{\Omega'}-\dfrac{1}{\Delta\phi}(e^{{I}^{+}E_i\Delta\phi} - e^{-{I}^{-}E_i\Delta\phi})\right]}.
\end{equation}

If $O = 0.5$, the real parts of the denominator and the energy-related terms of the numerator exactly cancel each other out. The resonance condition corresponds to the vanishing imaginary part of the denominator, yielding the point where $\tilde{p}_s(1)$ is maximized:

\begin{equation}\label{eq12}
e^{{I}^{+}E_i^*\Delta\phi} - e^{-{I}^{-}E_i^*\Delta\phi} = \frac{\omega\Delta\phi}{\Omega'}.
\end{equation}

For the specific case $O=0.5$ and $I=1$ discussed in the main text:

\begin{equation}\label{eq13}
\tilde{p}_s(1) \approx \frac{E_o}{2+i4\dfrac{\omega\Delta\phi-\Omega'(e^{E_i\Delta\phi}-1)}{\Omega'\Delta\phi(e^{E_i\Delta\phi}+1)}}.
\end{equation}

Although the modulus of the order parameter changes with increasing $E_o$, the phase condition for resonance remains largely unaffected. The deviation becomes significant only when $N$ is very small, as shown in Fig.~2 of the main text.


\section{External Potential Partition ($O \neq 0.5$)}

When the partition parameter $O \neq 0.5$, the cancellation between the real parts of the denominator and the numerator in Eq.~(\ref{eq11}) no longer holds. However, the resonance condition derived previously remains highly accurate for large $N$ and exhibits only minor deviations for very small $N$ (Fig.~S2).

In this regime, the optimal steady-state amplitude at resonance becomes:
\begin{equation}\label{eq13}
\left<R\right> \approx \tilde{p}_s(1) \approx E_o \frac{({O}^{+}e^{{I}^{+}E_i\Delta\phi} + {O}^{-}e^{-{I}^{-}E_i\Delta\phi})}{\left(e^{{I}^{+}E_i\Delta\phi} + e^{-{I}^{-}E_i\Delta\phi}\right)}.
\end{equation}

The optimal amplitude is independent of the phase step $\Delta\phi$ only in the symmetric case ($O = 0.5$). It is evident that the amplitude correlates positively with $\Delta\phi$ for $O > 0.5$, and negatively for $O < 0.5$. In the continuous limit ($\Delta\phi \to 0$), the optimal amplitude converges to $E_o/2$ regardless of $O$ (Fig.~S2).

This behavior has a clear physical interpretation. Driven by the intrinsic potential $E_i$, the forward reaction rate dominates ($k^+ > k^-$). When $O > 0.5$ (weighting the coupling energy towards the forward transition), the modulation factor $e^{E_o}$ acts more strongly on $k^+$, thereby enhancing the synchronization efficiency.

While the main text focuses on the symmetric case ($O = 0.5$), if $O$ is considered a tunable parameter, the optimal choice is evidently $O = 1$. In this scenario, a smaller $\Delta\phi$ becomes advantageous, which further corroborates the conclusions regarding the optimal state number presented in the main text.


\section{Mode Interactions from Higher-Order Terms}

Reviewing the derivation in the preceding section, we observe that the nearest-neighbor interaction terms arise specifically from the $\delta_{f,1}$ and $\delta_{f,N-1}$ components of the transformed rates $\hat{k}^\pm(f)$ via the convolution operation [Eq.~(\ref{eq6}) and (\ref{eq7})]. These specific components are introduced during the Fourier transform by the linear term $\mp O^{\pm}E_o \sin(\phi-\omega t)\Delta\phi$ in the expansion of the transition rates. 

To generate next-nearest-neighbor interaction terms (coupling $f$ with $f\pm 2$), the convolution would require contributions from the $\delta_{f,2}$ and $\delta_{f,N-2}$ terms of $\hat{k}^\pm(f)$. Physically, this implies that the expansion of $k^\pm(n)$ must include higher-frequency harmonics, such as $\sin(2\phi)$, which only appear in a second-order Taylor expansion. Generalizing this observation, we conclude that the $n$-th range mode interactions scale as the $n$-th order of the small parameter $O^{\pm}E_o\Delta\phi$ defined in Eq.~(\ref{eq5}). The derivation of mode equations involving second-order expansions is algebraically extensive and does not fundamentally alter the core conclusions; thus, we omit it here for brevity.


\section{Approximate Solution of Phase Vibration with Large $E_o$ in Resonance}

At resonance, the influence of the external signal $\omega$ and the internal drive $E_i$ becomes negligible compared to the strong coupling. Consequently, the probability distribution can be approximated by the Boltzmann distribution $P(\phi_n,t) \propto \exp(-U(\phi_n,t))$. Furthermore, for large $E_o$, the potential energy differences defined by $U(\phi_n,t)=-E_o\cos(\phi_n-\omega t)$ become significant, confining the system primarily to the states minimizing the potential. This allows us to restrict our analysis to the two states, $\phi_n$ and $\phi_{n+1}$, closest to the instantaneous signal phase $\omega t$.

Consider the interval where $\omega t$ lies between $\phi_n$ and $\phi_{n+1}$. We define the deviation from the midpoint as $\delta\phi=\omega t-(\phi_n+ \phi_{n+1})/2$. The phase difference between the signal and the collective phase $\Psi$ is given by a function $f(\delta\phi)$. The phase vibration amplitude corresponds to the variation range of this function:

\begin{equation}
f(\delta\phi) = \frac{\frac{\pi}{N}\exp(-U_1) - \frac{\pi}{N}\exp(-U_2)}{\exp(-U_1) + \exp(-U_2)} - \delta \phi,
\end{equation}

where the potentials for the two states are:

\begin{equation}
U_1 = -E_0 \cos\left(\frac{\pi}{N} - \delta \phi \right) \quad \text{and} \quad U_2 = -E_0 \cos\left(\frac{\pi}{N} + \delta \phi \right).
\end{equation}

The first term of $f(\delta \phi)$ can be rewritten using the hyperbolic tangent function:

\begin{equation}
f(\delta \phi) = \frac{\pi}{N} \tanh\left(\frac{U_2 - U_1}{2}\right) - \delta \phi.
\end{equation}

Substituting the expression for the potential difference, $U_2 - U_1$, and defining the stiffness parameter $A = E_0 \sin(\pi/N)$, we obtain:

\begin{equation}\label{fdletaphi}
f(\delta \phi) = \frac{\pi}{N} \tanh(A \sin(\delta \phi)) - \delta \phi.
\end{equation}

As $\delta \phi$ varies effectively within $[-\pi/N, \pi/N]$, $f(\delta \phi)$ exhibits periodic behavior. To determine the amplitude of this phase vibration, we locate the extrema by setting the derivative $f'(\delta \phi)$ to zero:

\begin{equation}
f'(\delta \phi) = \frac{\pi}{N} \cdot \text{sech}^2(A \sin(\delta \phi)) \cdot (A \cos(\delta \phi)) - 1 = 0.
\end{equation}

The extrema occur at small $\delta \phi$. Applying the small-angle approximations $\cos(\delta \phi) \approx 1$ and $\sin(\delta \phi) \approx \delta \phi$, the condition simplifies to:

\begin{equation}
\frac{\pi A}{N} \text{sech}^2(A \cdot \delta \phi) \approx 1 \implies \text{sech}^2(A \cdot \delta \phi) \approx \frac{N}{\pi A}.
\end{equation}

Assuming that the argument $A \cdot \delta \phi$ is sufficiently large, we employ the asymptotic approximation $\text{sech}(u) \approx 2e^{-u}$ to solve for $\delta \phi$:

\begin{equation}
\delta \phi_{\text{ext}} \approx \pm \frac{1}{2A} \ln\left(\frac{4 \pi A}{N}\right).
\end{equation}

The maximum occurs at $\delta \phi_{max} \approx + \frac{1}{2A} \ln\left(\frac{4 \pi A}{N}\right)$. Substituting $\delta \phi_{max}$ back into Eq.~(\ref{fdletaphi}), using $\sin(\delta \phi_{max}) \approx \delta \phi_{max}$, and applying the asymptotic expansion $\tanh(u) \approx 1 - 2e^{-2u}$ for large arguments, we find:

\begin{equation}
 f_{max} \approx \frac{\pi}{N} - \frac{1 + \ln(4\pi A / N)}{2A},
\end{equation}

where $A = E_0 \sin(\pi/N)$. By symmetry, $f_{min} = -f_{max}$. Thus, the total phase vibration amplitude is given by $\Delta\Psi=2f_{max} \approx \frac{2\pi}{N} - \frac{1 + \ln(4\pi A / N)}{A}$.


\section{Anomalous Dissipation at $N=3$ and Asymptotic Behavior in the Continuous Limit}

Figure~3(b) in the main text exhibits an anomalous increase in dissipation $\dot{W}$ for $N=3$ compared to larger $N$. This arises from high discreteness: states cannot perfectly align with the external signal phase $\omega t$. Consequently, clocks are forced away from the potential minima, and transitions involve potential energy differences significantly larger than the intrinsic drive $E_i\Delta\phi$. This suggests that extremely small state numbers are energetically unfavorable for precise timekeeping, reinforcing our conclusions on the optimal choice of $N$.

This framework also clarifies the asymptotic behavior in Fig.~1(d). In the continuous limit with large $E_o$, the probability distribution becomes tightly confined around the potential minimum ($\omega t$). The external potential compensates for any mismatch between the intrinsic and external frequencies. The energy cost to maintain this phase-locked state asymptotically approaches the resonant dissipation $\dot{W}_{res} = \omega E_i^*$, regardless of the detuning parameter $\alpha$.

A similar mechanism applies to the discrete case. Strong coupling $E_o$ maintains the collective amplitude near resonance over a wider range of $E_i$, implying a flatter response (smaller second derivative of $\langle R \rangle$ about $E_i$), which enhances parameter stability (Fig.~S1). Note that for very small $N$, the optimal drive $E_i^*$ deviates significantly from theoretical predictions due to the breakdown of approximations. This is the reason why we did not display this data point in Fig.~3(b). This does not affect our conclusion.

\begin{figure*}[t]\centering
\includegraphics[width=1\linewidth]{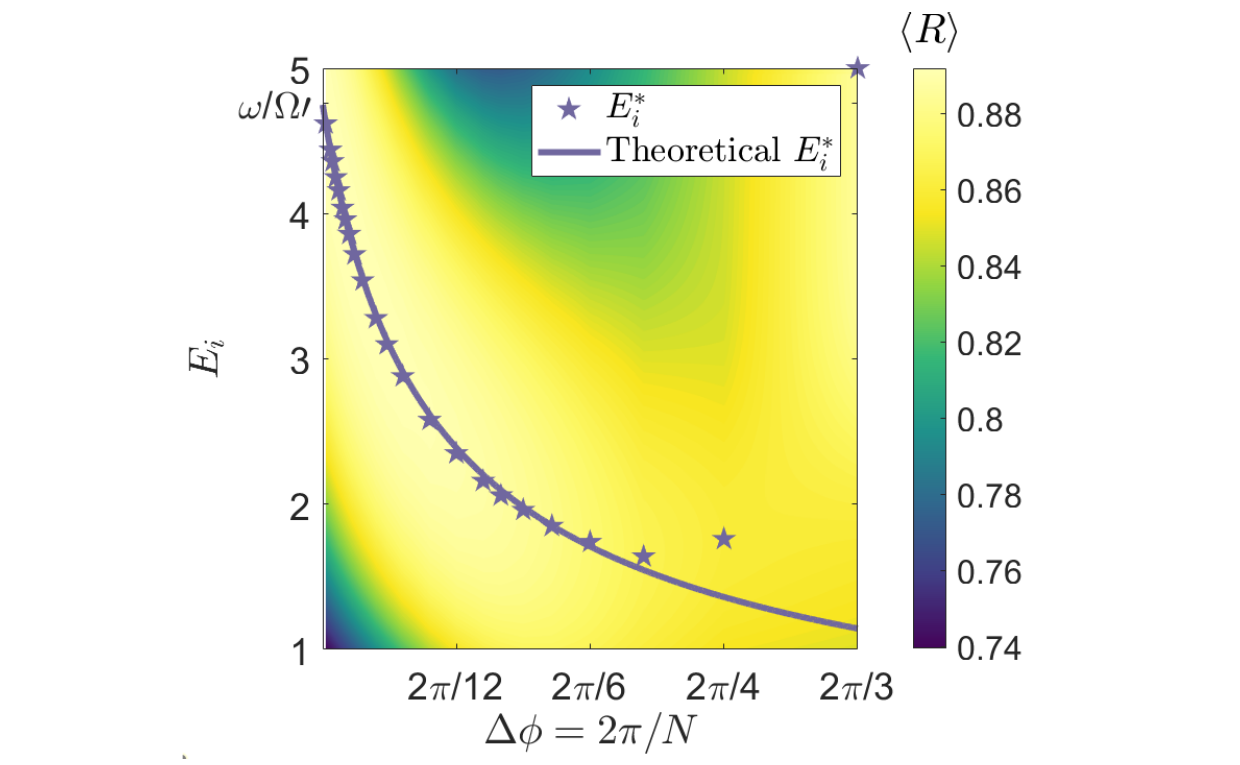}\\
\caption{ Same to Fig. 2(a) in the main text. $E_o=5, \Omega'=0.055, \omega=2\pi/24$.}\label{figs1}
\end{figure*}

\begin{figure*}[t]\centering
\includegraphics[width=1\linewidth]{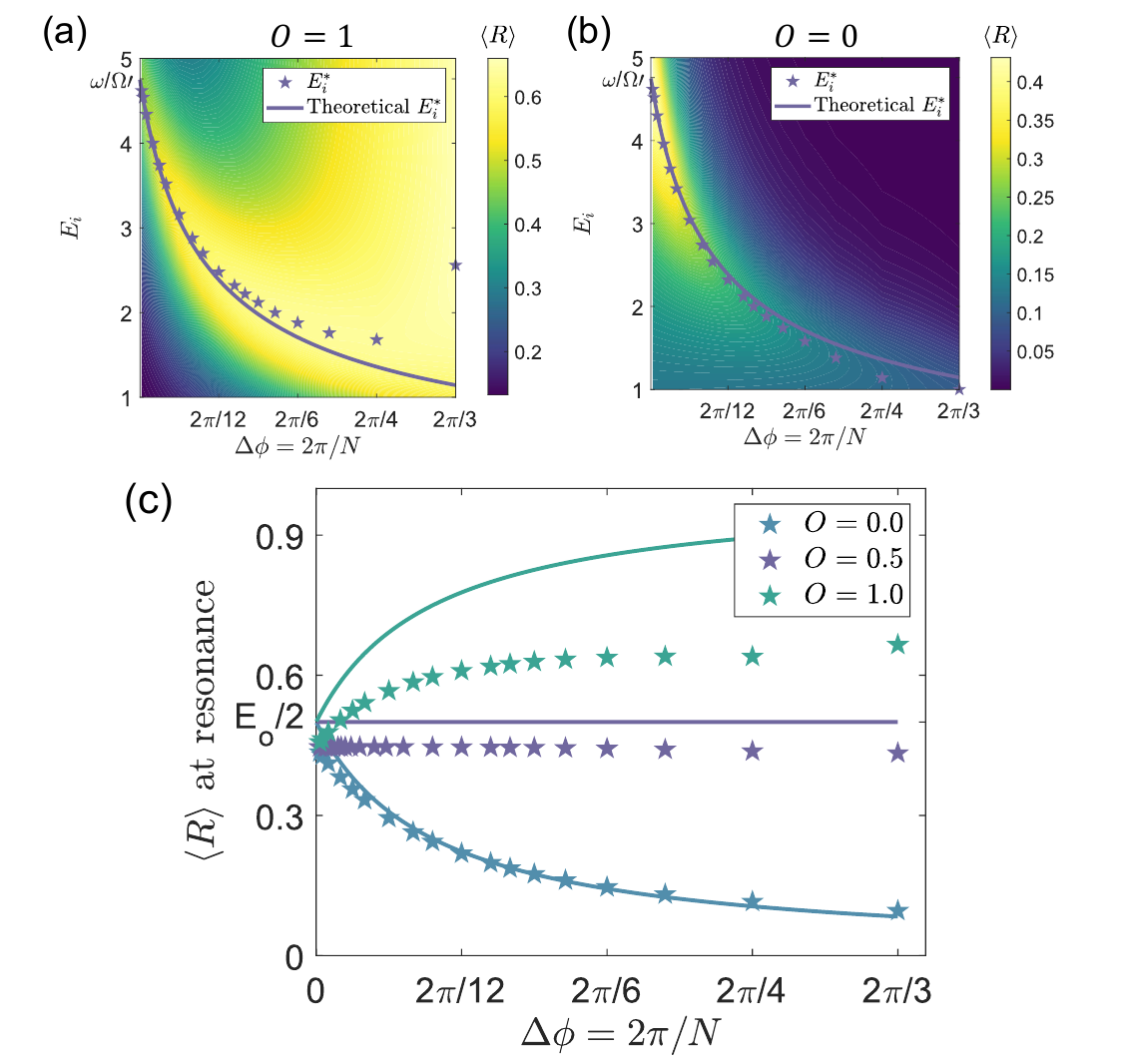}\\
\caption{ (a) and (b) show the same to Fig. 2(a) in the main text. $E_o=1, \Omega'=0.055, \omega=2\pi/24$. (a)$O=1$.  (b)$O=0$. (c)The $\left<R\right>$ at resonance in different $\Delta \phi$ and $O$. The solid curves represent the theoretical predictions.}\label{figs2}
\end{figure*}
